\documentclass[a4paper,11pt]{article}
\usepackage{latexsym}
\usepackage{epsfig, subfigure, subfloat, graphicx, float }
\usepackage{amssymb}
\usepackage{amsmath}
\usepackage{epstopdf}
\usepackage{authblk}
\newcommand*{\mydprime}{^{\prime\prime}\mkern-1.2mu}

\title{\bf Thawing in a coupled quintessence model}  \author[1]{M. Honardoost \thanks{m\_honardoost@sbu.ac.ir}}
\author[2]{H. Mohseni Sadjadi\thanks{mohsenisad@ut.ac.ir}}
\author[1]{H. R. Sepangi \thanks{hr-sepangi@sbu.ac.ir}}
\affil[1]{\small Department of Physics, Shahid Beheshti University, G.C., Evin, Tehran
19839, Iran}
\affil[2]{\small Department of Physics, University of Tehran, P. O. B. 14395-547, Tehran 14399-55961, Iran}

\begin{document}

\maketitle
\begin{abstract}
We consider the thawing model in the framework of coupled quintessence model. The effective potential has $Z_2$ symmetry which is broken spontaneously when the dark matter density becomes less than a critical value leading the quintessence equation of state parameter to deviate from -1. Conditions required for this procedure are obtained and analytical solution for the equation of state parameter is derived.
\end{abstract}

\section{Introduction}

Current observational evidences indicate that our universe,  consists of nearly 68\% dark energy, the mysterious energy component with negative pressure, and nearly 27\% of an unusual matter component called dark matter \cite{Planck}. Although , considering dark energy as the cosmological constant $\Lambda$ was proposed in standard $\Lambda$ cold dark matter, $\Lambda \mathcal{CDM}$ model \cite{Weinberg}, due to problems such as the coincidence and fine tuning, it may be more interesting to think about dark energy as a dynamical component. For this purpose, dynamical scalar fields $(\phi(t))$ in the framework of quintessence models are suitable candidates, which have been widely studied before \cite{quint}.
In $\Lambda\mathcal{CDM}$ model, the equation of state (EoS) parameter of dark energy is exactly $-1$ while observations show the EoS parameter of dark energy, $w_\phi$, satisfies $-0.9\leq w_\phi \leq-1.1 $ \cite{Planck}, which means that $w_\phi$  may not be exactly -1.
One possible way to explain this is to consider dark energy as ``Thawing quintessence'' \cite{linder}. Thawing scalar fields hardly move in the past but roll down the potential recently. These type of models are studied in detail in \cite{chiba} {\it but without considering any interaction between dark sectors}.
To alleviate the coincidence problem, various interactions between dark matter and dark energy have been proposed in the literature \cite{int}.
In our study we consider {\it coupled quintessence thawing model. We use this coupling to explain deviation of $w_\phi$ from $-1$ through  spontaneous symmetry breaking of the effective potential}.

  In our investigation we consider a coupled dark energy model in which the effective potential of the scalar field depends on dark matter density such that the scalar field resides in the minimum of the effective potential during the time when the density of  matter is greater than a critical value \cite{kh}. However, when the density becomes less than a critical density, the shape of the potential changes and, as we shall see, the initial symmetry spontaneously breaks and the scalar field rolls down to a new minimum and the equation of state of $\phi$ deviates from -1. {\it Unlike \cite{kh} we do not restrict ourselves to a specific potential or coupling and do not consider any additional coupling between dark energy and baryonic matter. So we relax the local gravitational tests and the gravitational screening effect which pose a large mass constraint on the quintessence mass.} As the nature of dark sectors is not specified, considering such phenomenological interactions between (only) dark matter and dark energy is customary in the literature.

{\it  Our main goal is to derive an analytic expression for the equation of state parameter of the quintessence in terms of the scale factor near the maximum of the potential and study analytically the deviation of $w_\phi$ from $-1$ after the symmetry breaking}. This is an extension of studies on thawing models in the literature where such an interaction was not considered \cite{chiba} and the reason behind the scalar field leaving its initial position was not clearly shown \cite{scher}.

The plan of the paper is as follows:
In the second section,  we discuss some preliminaries regarding our coupled quintessence. In the third section, to establish our model, we propose the required propositions  which are necessary to have the desired symmetry breaking. In the Sec.4, {\it which contains our main results}, we  study analytically the evolution of the scalar field and its EoS parameter near the local extremum of the potential where $ w_\phi\simeq-1$  and plot the EoS parameter of dark energy numerically.
We use units $\hbar=c=8\pi G=1$ through the paper.

\section{Preliminaries}
We consider a universe described by a spatially flat Friedmann–-Lemaître-–Robertson–-Walker (FLRW) metric, filled by a quintessence scalar field dark energy $\phi$, cold (pressureless) dark matter $\rho_d$, and baryonic matter $\rho_b$.
The dark energy is coupled to dark matter via a source term, $f(\phi)\dot{\phi}\rho_d$ \cite{amendola}. For example One can see such  a coupling after a conformal transformation of Brans-Dicke theory or in string theory \cite{amendola1}. We consider the interaction only between dark sectors and the baryonic matter does not couple to dark  sectors

\begin{eqnarray}\label{1}
&&\ddot{\phi}+3H\dot{\phi}+V_{,\phi}=-f(\phi)\rho_d\nonumber \\
&&\dot{\rho_d}+3H\rho_d=f(\phi)\dot{\phi}\rho_d\nonumber  \\
&&\dot{\rho_b}+3H\rho_b=0.
\end{eqnarray}
The Friedmann equations are
\begin{equation}\label{2}
H^2={1\over 3}\left({1\over 2}\dot{\phi}^2+V(\phi)+\rho_b+\rho_d\right),
\end{equation}
and
\begin{equation}\label{3}
\dot{H}=-{1\over 2}\left(\dot{\phi}^2+\rho_b+\rho_d\right).
\end{equation}
Note that (\ref{1}), and only one of the two equations (\ref{2}) and (\ref{3}) can be considered as independent equations. $H={\dot{a}\over a}$ is the Hubble parameter and $a(t)$ is the scale factor. By convention, we take  $a=1$ at the present time.
The equations of motion can also be obtained via the action
\cite{trodden}
\begin{equation}\label{t}
S=\int d^4x\sqrt{-g}\left(\frac{1}{2}R-{1\over 2}g^{\mu \nu}\partial_\mu \phi \partial_\nu \phi-V(\phi)\right)+S_m[\tilde{g^i}_{\mu \nu},\psi^i],
\end{equation}
where $R$ is the scalar curvature, $\phi$ is the quintessence and $S_m$ is the action corresponding to other species $\psi^i$ . The quintessence is coupled to $\psi^i$ via $\tilde{g^i}_{\mu \nu}=A_i^2(\phi)g_{\mu \nu}$, where $A_i(\phi)$ is a positive function. We take different couplings for different species \cite{trodden}. As we aim to consider interaction only between dark sectors, we assume that $A_i$ is nontrivial $A_i\neq 1$ only for dark species. If for dark matter we take
\begin{equation}
A_i=A(\phi)=\exp\left(\int f(\phi) d\phi\right),
\end{equation}
by variation of the action we get the equations of motion (\ref{1}) and Friedmann equation (\ref{2}).

By introducing the density, $\rho$,
\begin{equation}\label{4}
\rho_d=A(\phi)\rho,
\end{equation}
we find that
\begin{equation}\label{5}
\dot{\rho}+3H\rho=0,
\end{equation}
whose solution is given simply by $\rho=\rho_0a^{-3}$, where $\rho(a=1)\equiv \rho_0$.
This $\rho$ is not the physical density but a mathematical object which help us to solve and interpret some of our equations.
The quintessence equation of motion can be rewritten as
\begin{equation}\label{6}
\ddot{\phi}+3H\dot{\phi}+V_{,\phi}=-A_{,\phi}\rho.
\end{equation}
Hence in the evolution of $\phi$, $V^{eff}(\phi,\rho)$ defined by $V^{eff}_{,\phi}\equiv V_{,\phi}+A_{,\phi}\rho$, plays the role of an effective action.
\begin{equation}\label{7}
\ddot{\phi}+3H\dot{\phi}+V^{eff}_{,\phi}=0.
\end{equation}

{\it It is worth noting that $V^{eff}(\phi,\rho)$ depends on $\phi$ as well as on $\rho$. Therefore the value of $\phi$ which extremizes the effective potential,  is not a constant generally and may be $\rho$ (and consequently time) dependent.}

\section{Conditions required for thawing via symmetry breaking}

In this part, necessary conditions for $A(\phi)$ and $V(\phi)$ allowing the thawing process via symmetry breaking are investigated. Usually in the literature, some specific examples for $V(\phi)$ and $A(\phi)$ are used to study the symmetry breaking in symmetron models \cite{kh}, but we try to discuss the subject generally.

To attribute the evolution of $w(\phi)$ in thawing model to spontaneous symmetry breaking in our coupled quintessence model,  following \cite{kh}, we assume that the effective potential has a $Z_2$ symmetry which is broken by the evolution of $\rho$. We consider two following eras:\newline

I- When $\rho$ is greater than a critical value, the quintessence field stays at the minimum of its effective potential. \vspace{1cm}

II- When $\rho$ becomes less than a critical value (specified by the parameters of the model)the shape of effective potential changes and the $Z_2$ symmetry is spontaneously broken.
The previous minimum becomes an unstable point, and the scalar field moves towards the new minimum of the effective potential. \newline

To study the condition (I), let us investigate whether the scalar field can reside at the minimum of the effective potential $V^{eff}(\phi,\rho(t))$, in the sense that
\begin{equation}\label{8}
V_{,\phi}+A_{,\phi}\rho=0.
\end{equation}
As $\rho=\rho_0a^{-3}$, let us see whether (\ref{8}) gives a constant or time dependent solution for $\phi$.
Note that since we consider (\ref{8}) as an exact equation, in contrast to works where (\ref{8}) is only considered as an approximation, using the equation of motion (\ref{6}), we immediately find
\begin{equation}\label{9}
\ddot{\phi}+3H\dot{\phi}=0.
\end{equation}
We consider two possible situations (i and ii) and show that the first one (i.e. i) is not acceptable.\noindent \\
i- $A_{,\phi}\neq 0$ at the minimum of the effective potential:
In this situation (\ref{8}) determines $\rho$ in terms of $\phi$,
\begin{equation}\label{10}
\rho=-{V_{,\phi}\over A_{,\phi}}\equiv h(\phi),
\end{equation}
where $h(\phi)$ is an analytical function of $\phi$. The solution of (\ref{9}) is
\begin{equation}\label{11}
\dot{\phi}=Ca^{-3},
\end{equation}
where $C$ is a numerical constant. Recalling that $\rho$ satisfies (\ref{5}) and by using  (\ref{10}) and (\ref{11}),
we can obtain an integral equation for the scale factor
\begin{equation}\label{12}
h(\int Ca^{-3} dt)=\rho_0 a^{-3}.
\end{equation}
But this is not the whole story: {\it We have one more equation to be satisfied, i.e. the Friedmann equation (\ref{2}) or (\ref{3})}.
As the Friedmann equation is independent of the equations (\ref{5}), (\ref{7}), it is natural to expect that it may be not consistent with (\ref{12}).
To clarify this issue, let us consider,  as an example, a quintessence field with the potential
\begin{equation}\label{13}
V(\phi)=-{1\over 2}\mu^2\phi^2+{\lambda\over 4}\phi^4+v,
\end{equation}
where $\lambda$ and $v$ are two positive constants , and
\begin{equation}\label{14}
A(\phi)=1+{M^2\over 2}\phi^2.
\end{equation}
Note that $\dot{\rho}=h_{,\phi}\dot{\phi}$. Using (\ref{5}), this leads to  $-3H\rho=h_{,\phi}\dot{\phi}$, whose time derivative gives
 \begin{equation}\label{15}
 -3\dot{H}\rho-3H\dot{\rho}=h_{,\phi\phi}{\dot\phi}^2+h_{,\phi}\ddot{\phi}.
 \end{equation}
 But from $\dot{\rho}=h_{,\phi}\dot{\phi}$ we have $ -3H\dot{\rho}=-3H h_{,\phi}\dot{\phi}$, which by using (\ref{9}) gives
 \begin{equation}\label{rr1}
 -3H\dot{\rho}=h_{,\phi}\ddot{\phi}.
 \end{equation}
Now, using (\ref{rr1}), (\ref{15}) reduces to
 \begin{equation}\label{16}
 -3\dot{H}\rho=h_{,\phi\phi}\dot{\phi}^2.
 \end{equation}
For (\ref{13}) and (\ref{14}),  $h_{,\phi\phi}=\left({\mu^2\over M^2}-{\lambda \phi^2\over M^2}\right)_{,\phi \phi}=-{2\lambda \over M^2}<0$, therefore from (\ref{16}) we find that $\dot{H}>0$. This is in contradiction with the equation of motion (\ref{2})
\begin{equation}
\dot{H}=-{1\over 2}\left(\dot{\phi}^2+\rho_b+ A(\phi)\rho\right),
\end{equation}
 which dictates that $\dot{H}$ is negative in our model (note that the quintessence is assumed to be real and $A(\phi)>0$). Despite this, note that in the presence of interaction it is possible to obtain an effective equation of state parameter for the dark sector such that $w^{eff.}_\phi<-1$  \cite{effective}.
In the same way, it is not hard to see that the assumption (i) is not consistent with the Chameleon model \cite{khoury2}.\noindent\vspace{3mm} \\
ii- $A_{,\phi}=0$. However if we assume that at the minimum of the effective potential,  $A_{,\phi}=0$, then (\ref{8}) does not put any constraint on $\rho$ nor on $a$ (note that $\rho\neq 0$). As $A$ is a specific function of $\phi$, $A_{,\phi}=0$ fixes the value of $\phi$, i.e $\phi=\phi_c$, where $\phi_c$ is a time independent constant.  Hence $C=0$ in (\ref{11}).  Besides,  from (\ref{8}), we conclude $V_{,\phi}(\phi_c)=0$.  In this case, the potential of the scalar field plays the role of a cosmological constant and the Friedmann equation holds consistently. For example,  for the model specified by (\ref{13}) and (\ref{14}), the quintessence may lie at $\phi=\phi_c=0$, initially, where  $A_{,\phi}(0)=0$, but for Chameleon model with $A(\phi)=\exp(\beta \phi)$, the quintessence can not reside at the minimum of the effective potential in any situations.

{\it Whence at the stage (I)(mentioned in the beginning of the section), $\phi$ must be a constant: $\phi=\phi_c$, $\dot{\phi}=0$, and $A_{,\phi}(\phi_c)=V_{,\phi}(\phi_c)=0$}. In this era the EoS parameter of the quintessence is
\begin{equation}\label{17}
w_\phi={{1\over 2}\dot{\phi}^2-V(\phi)\over {1\over 2}\dot{\phi}^2+V(\phi)}=-1,
\end{equation}
as required.

{\it{It is important to note that (I), which states that the scalar field resides exactly at the the minimum of the effective potential,  is much more restrictive than the adiabaticity situation used in the literature for models such as Chameleon,  where the field traces the minimum of the effective potential \cite{khoury2} and (\ref{8}) is only an approximation and so (\ref{9}) does not hold and higher derivatives of the effective potential should be considered.}}

In the first stage, i.e. stage I, as $V^{eff}_{,\phi}\big|_{\phi=\phi_c}=0$, then  $\ddot{\phi}+3H\dot{\phi}+V^{eff}_{,\phi}=0$ has a constant solution $\phi=\phi_c$. In this stage $\rho_d$ and $\rho_b$ evolve similarly; $\rho_d \propto \rho_b\propto a^{-3}$. To see whether the system is stable in the $\phi$ direction, i.e. small quantum fluctuations do not force $\phi$ to leave $\phi_c$, it is sufficient to consider second derivative of the effective potential. As long as $m_{eff.}^2=\left(V_{,\phi\phi}+A_{,\phi\phi}\rho\right)\big|_{\phi=\phi_c}>0$ (the shape of the effective potential is convex), $\phi$ continues to stay classically at $\phi=\phi_c$. In this stage the vacuum expectation value of the scalar field is  $\left<\phi\right>=\phi_c$. But during this epoch $\rho$  decreases, and if we establish our model such
that $V_{,\phi \phi}(\phi_c)<0$ and $A_{,\phi \phi}(\phi_c)>0$, we eventually have  $m_{eff.}^2\big|_{\phi=\phi_c}<0$. At this moment, the scalar field becomes tachyonic and $\phi_c$ is no more the true vacuum,  and $\phi$ recedes from $\phi_c$ and moves towards the true vacuum, $\phi_{c1}$. Finally we will have $<\phi>=\phi_{c1}$ after a time of order $m_{eff.}^{-1}$. In this stage $\dot{\phi}\neq 0$ and the EoS parameter deviates from $w_\phi =-1$, due to the mentioned symmetry breaking.

To see, via a dynamical system analysis which includes the late time behavior of all ingredients of the system,  that
the quintessence cannot stay at the initial point forever (playing the role of a cosmological constant), one can also use a simple and brief phase space analysis. Note that in this brief discussion we only consider the late time evolution,  and the (transient) stability of the scalar field which occurs at the first stage (I) and studied above, is not discussed in this dynamical analysis.

By defining dimensionless variables \cite{auto}
$x={\dot{\phi}\over \sqrt{6}H},\,\,y={\sqrt{V}\over \sqrt{3}H},\,\,z_{d}={\sqrt{\rho_d}\over \sqrt{3}H},\,\,z_{b}={\sqrt{\rho_b}\over \sqrt{3}H}$, we find
\begin{eqnarray}\label{18}
{dx\over dN}&=&-3x-\sqrt{3\over 2}fy^2+3x^3+{3\over 2}z_b^2x+{3\over 2}z_d^2x-\sqrt{3\over 2}z_d^2g,\nonumber \\
{dz_d\over dN}&=&\sqrt{6}xz_dg+3z_dx^2-{3\over 2}z_d+{3\over 2}z_d^3+{3\over 2}z_dz_b^2,\nonumber \\
{dz_b\over dN}&=&3x^2z_b-{3\over 2}z_b+{3\over 2}z_b^3+{3\over 2}z_bz_d^2,\nonumber \\
{df\over dN}&=&\sqrt{6}xf^2\left(\Gamma-1\right),\nonumber \\
{dg\over dN}&=&\sqrt{6}xg^2\left(\zeta-1\right),
\end{eqnarray}
where $N\equiv\ln a$, $f\equiv {V'\over V}$, $g\equiv {A'\over A}$, $\Gamma\equiv {VV\mydprime\over V'^2}$, $\zeta\equiv {AA\mydprime\over A'^2}$.
From (\ref{2}), we have $x^2+y^2+z_d^2+z_b^2=1$, so $y$ is not considered as an independent variable and may be rewritten in terms of other variables. Therefore (\ref{18}) becomes an autonomous system of differential equations by taking $\Gamma=\Gamma(f,g)$ and $\zeta(f,g)$. The point  $D\equiv \{\bar{x}=0,\bar{z_d}=0,\bar{z_b}=0\}, \bar{f}=f(\bar{\phi})=0$ is a critical point of the system, where the energy of the Universe is coming only from the scalar field potential, at late times. This point is exactly the same point of the phase space if the scalar field could stay at $\phi=\phi_c$. Note that the relations $\bar{z_d}=0,\bar{z_b}=0$ hold only at late times and are not true in the first stage where $V^{eff}_{,\phi \phi}\big|_{\phi=\phi_c}>0$.
However this can  happen provided that $D$ is a stable point. To study this stability let us consider small perturbations around $D$, i.e.  $x=\bar{x}+\delta x,\,\, z_d=\bar{z}_d+\delta z_d,\,\,  z_b=\bar{z}_b+\delta z_b,\,\,   g=\bar{g}+\delta g, \,\,f=\bar{f}+\delta{f}$. Consider matrix $M$ defined by
\begin{equation}\label{19}
{d\over dN}{ \left( \begin{array}{ccc}
\delta x \\
\delta z_d \\
\delta z_b\\
\delta f\\
\delta g
\end{array} \right)}=M \left( \begin{array}{ccc}
\delta x \\
\delta z_d \\
\delta z_b\\
\delta f\\
\delta g
\end{array} \right).
\end{equation}
Hence at the critical point $D$
\begin{equation}\label{20}
M= \left( \begin{array}{ccccc}
-3 &0 & 0 &-{\sqrt{6}\over 2}& 0 \\
0 & -{3\over 2} & 0 & 0 & 0 \\
0 & 0 & -{3\over 2} & 0 & 0 \\
\sqrt{6}{V\mydprime(\bar{\phi}) \over V(\bar{\phi})}& 0 & 0&0&0\\
\sqrt{6} \bar{g}^2\left(\zeta(0,\bar{g})-1\right)& 0 & 0&0&0
\end{array} \right).
\end{equation}
Eigenvalues of $M$ are $\lambda=0,\lambda=-{3\over 2}, \lambda=-{3\over 2}, \lambda=-{3\over 2}\pm {1\over 2}\sqrt{9-12{V\mydprime(\bar{\phi}) \over V(\bar{\phi})}}$.
Assuming $V(\bar{\phi})>0$, which is necessary to have a real Hubble parameter, one obtains a positive eigenvalue when $ V\mydprime(\bar{\phi})<0$, indicating the instability at the point $D$ where the potential is concave (note that at point D where $\rho_d$ and $\rho_b$ tend to zero, the effective potential becomes asymptotically the same as the potential at late time).

Note also that condition $V\mydprime(\bar{\phi})<0$, which leads to a positive eigenvalue i.e.  $\lambda=-{3\over 2}+{1\over 2}\sqrt{9-12{V\mydprime(\bar{\phi}) \over V(\bar{\phi})}}>0$, is a sufficient condition for having instability at $\phi=\phi_c$ at the second stage. As we require that at late time the scalar field  becomes unstable at $\phi=\phi_c$, we adopt $V\mydprime(\bar{\phi})<0$ for our proposed model. However this is not a necessary condition for occurrence of instability;  as one of the eigenvalues is zero, the critical point is non-hyperbolic and even if the other eigenvalues are negative, one must apply other tests such as applying center manifold theorem to check the stability or instability of the system at the critical point \cite{rev}. So it may be possible that the critical point is unstable at late times  even for  $V\mydprime(\bar{\phi})>0$.

In summary,  to satisfy our proposal II, we assume that the scalar field stays at the maximum of its potential (where $V'=0$, $V \mydprime<0$, and $V^{eff}_{,\phi, \phi}>0$) initially. This is the minimum of the effective potential and this stay is transient.  Later,  when this point is no longer the minimum of the effective potential, due to instabilities, the scalar field moves and we obtain a dynamical field, that is $\dot{\phi}\neq 0$. From (\ref{17}), it is clear that
$w_\phi>-1$.

Theories like (\ref{t}), with coupling between dark matter and the quintessence may exhibit instability in the adiabatic regime where the quintessence follows the minimum of its effective potential \cite{afshordi}. The sound speed squared becomes negative and  perturbations grow exponentially.
In \cite{trodden}, by considering the action (\ref{t}), it was shown that such instability occurs only for coupling satisfying
\begin{equation}\label{c}
{d\ln(A(\phi))\over d\phi}\gg {1\over M_P},
\end{equation}
where $M_p$ is the reduced planck mass (we have taken $M_p=1$ in the units used in the paper).  So one can evade the instability by choosing appropriate parameters.
For example if one chooses $A(\phi)=1+{\phi^2\over M^2}+\mathcal{O}\left({\phi^4\over M^4}\right)$, where $M$ is a mass scale such that ${\phi^2\over M^2}\ll 1$, the condition (\ref{c}) reduces to $\phi\gg {M^2\over M_p}$.  So for  $\phi< {M^2\over M_p}$ we can evade this instability.

Summarising, the model (\ref{1}) can describe the thawing model via $Z_2$ symmetry breaking. For this purpose, it is sufficient to take an even potential and even $A(\phi)$ which at the initial point $\phi_c$ satisfy $V_{,\phi}(\phi_c)=0$,
$A_{,\phi}(\phi_c)=0$, $V_{,\phi\phi}(\phi_c)<0$, and $A_{,\phi\phi}(\phi_c)>0$.

\section{Evolution of the model near the extremum of the potential}
In this section we begin by deriving an analytical expression for the EoS parameter of coupled dark energy. Although $\phi$ plays the role of a cosmological constant initially, afterwards, by the symmetry breaking, the scalar field rolls down from $\phi_c$ and $w_{\phi}$ deviates from $-1$.  The difference of this model and $\Lambda\mathcal{CDM}$ is that it describes a dynamical dark energy with $w>-1$ while for $\Lambda\mathcal{CDM}$ we have always $w=-1$. This may help to alleviate the coincidence problem and also improve the compatibility of the model with astrophysical data \cite{Gomez}.

In the following we study evolutions of the scalar field $\phi$ and $w_{\phi}$ near $\phi=\phi_c$,  when  $w_\phi\simeq -1$. We assume that the present epoch is in this regime and take $a_0=1$. By the subscript $0$, we denote the value of the corresponding parameter at present time. Note that our computation is only valid for the present epoch when assume that the scalar field is near $\phi_c$, that is $\phi\simeq \phi_c$. 
 In the absence of interaction, one obtains \cite{solution}
 \begin{equation}\label{21}
 H^2=H_0^2\left(\Omega_{\phi 0}+(1-\Omega_{\phi0})a^{-3}\right).
 \end{equation}
 Relative densities are defined by $\Omega_i\equiv {\rho_i\over 3H_0^2}$, e. g.   $\Omega_{\phi}\equiv {\rho_{\phi }\over 3H_0^2}$, where $\rho_\phi$ is the energy density of the scalar field. As $w_\phi\simeq -1$, we have $\dot{\rho_{\phi}}\simeq 0$ implying that $\Omega_{\phi}\simeq \Omega_{\phi 0}$, which was used in derivation of  (\ref{21}). The solution of (\ref{21}) is \cite{solution}
 \begin{equation}\label{22}
 a=\left(1-\Omega_{\phi 0}\over \Omega_{\phi 0}\right)^{1\over 3}\sinh^{2\over 3}\left(t\over t_\Lambda\right),
 \end{equation}
where $t_\Lambda={2\over \sqrt{3\rho_{\phi 0}}}$.

In our coupled quintessence model we have an interaction term between dark sectors
\begin{eqnarray}\label{23}
&&\dot{\rho}_d+3H\rho_d=A_{,\phi}A^{-1}\rho_d \sqrt{\gamma_\phi \rho_\phi}, \nonumber \\
&&\dot{\rho}_\phi+3H\gamma_\phi\rho_\phi=-A_{,\phi}A^{-1}\rho_d \sqrt{\gamma_\phi \rho_\phi},
\end{eqnarray}
where $\gamma_\phi=w_\phi+1$ and we have used $\dot{\phi}^2=\gamma_\phi \rho_\phi$. Note that in the regime under study $\gamma_\phi\simeq 0$.
The scalar field satisfies  equation (\ref{6}) which, by defining  $\phi-\phi_c=ua^{-3\over 2}$, can be rewritten as
\begin{equation}\label{24}
\ddot{u}+{3\over 4}Pu+a^{3\over 2}V^{eff}_{,\phi}=0.
\end{equation}
To derive this relation we have used $\left({\ddot a\over a}+{1\over 2}\left({\dot{a}\over a}\right)^2\right)=-{1\over 2}P$, where $P$ is the total pressure of the Universe. Using  $P_\phi\simeq -\rho_\phi\simeq -V(\phi_c)$,
and the relation
\begin{equation}\label{25}
V+A\rho\simeq V(\phi_c)+A(\phi_c)\rho+{1\over 2}\left(V_{,\phi\phi}(\phi_c)+A_{,\phi\phi}(\phi_c)\rho\right)\left(\phi-\phi_c\right)^2,
\end{equation}
which holds near the extremum of the potential $\phi_c$, we finally obtain
\begin{equation}\label{26}
\ddot{u}+\left(V_{,\phi \phi}(\phi_c)+A_{,\phi \phi}(\phi_c)\rho_0a^{-3} -{3\over 4}V(\phi_c)\right)u=0.
\end{equation}
Setting $A=1$, this reduces to the result obtained in \cite{scher} and \cite{solution}.
Now in the regime $\gamma_\phi\simeq 0$, we approximate $a(t)$ in (\ref{26}) by (\ref{22}).
Changing the variables $x\equiv \cosh(\tau)$ and $u(x)\equiv (x^2-1)^{1\over 4}y(x)$, where $\tau\equiv{t\over t_\Lambda}$ is a dimensionless time, (\ref{26}) reduces to the Legendre equation
\begin{equation}\label{27}
(1-x^2){d^2y\over dx^2}-2x{dy\over dx}+\left(\nu(\nu+1)-{\mu^2\over 1-x^2}\right)y(x)=0.
\end{equation}
In terms of the scale factor, the parameter $x$ is given by
\begin{equation}\label{28}
x=\sqrt{1+{\Omega_{\phi 0}\over \Omega_{m0}}a^3},
\end{equation}
where $\Omega_{m0}= {\rho_{m0}\over 3H}$ and $\rho_m$ is the sum of the dark matter and baryonic matter densities. Also
\begin{eqnarray}\label{29}
-\nu(\nu+1)&=&{4V\mydprime(\phi_c)\over 3 \rho_{\phi0}}-{V(\phi_c)\over \rho_{\phi 0}}+{1\over 4},\nonumber \\
\mu^2&=&{1\over 4}-{4A\mydprime(\phi_c)\rho_{d0}\over 3 A(\phi_c)\rho_{m0}}.
\end{eqnarray}
To get more physical intuition on $\mu$ and $\nu$, we use
\begin{equation}\label{30}
V\mydprime(\phi_c)+{A\mydprime(\phi_c)\over A(\phi_c)}\rho_{dc}=0,
\end{equation}
where $\rho_{dc}=\rho_d(a_c)$ is the value of dark matter density when the effective potential becomes convex. Using (\ref{30}), we rewrite (\ref{29}) as
\begin{eqnarray}\label{31}
-\nu(\nu+1)&=&\left({4\over 9\Omega_{\phi 0}}\right)\left({V\mydprime(\phi_c)\over H_0^2}\right)-{V(\phi_c)\over \rho_{\phi 0}}+{1\over 4}, \nonumber \\
\mu^2&=&\left({4a_c^3\over 9\Omega_{m0}}\right)\left({V\mydprime(\phi_c)\over H_0^2}\right)+{1\over 4},
\end{eqnarray}
with ${V\mydprime(\phi_c)\over H_0^2}\sim {-m_\phi^2\over H_0^2}$ where $m_\phi$ is the mass of $\phi$, e.g. for the potential (\ref{13}), $m_{\phi}=\sqrt{2}\mu$ which is the mass of excitations around the true minimum after the symmetry breaking. General real solution of the scalar field is given by \cite{Legendre}
\begin{equation}\label{32}
\phi=\phi_c+\left({\Omega_{\phi 0}\over \Omega_{m0}}\right)(x^2-1)^{-{1\over4}}\left(C_1 P^{-\mu}_\nu(x)+C_2 e^{-i\pi\mu}Q^\mu_\nu(x)\right).
\end{equation}
 $C_1$ and $C_2$ are two real constants and $P^{\mu}_{\nu}$ and $Q^{\mu}_{\nu}$ are associated Legendre functions of the first and second kind respectively.
Near $w=-1$, where $\dot{\phi}^2\ll V$, we have
\begin{equation}\label{33}
w_\phi={{1\over 2}\dot{\phi}^2-V(\phi)\over {1\over 2}\dot{\phi}^2+V(\phi)}\simeq -1 +{\dot{\phi}^2\over V(\phi_c)}.
\end{equation}
From ${d\phi\over dt}=\dot{a} {dx\over da}{d\phi\over dx}$, we derive
\begin{equation}\label{34}
{d\phi\over dt}={3\over 2}H_0{\Omega_{\phi 0}\over \sqrt{\Omega_{m0}}}\sqrt{x^2-1}{d\over dx}\left((x^2-1)^{-{1\over4}}\left(C_1 P^{-\mu}_\nu(x)+C_2 e^{-i\pi\mu}Q^\mu_\nu(x)\right)\right).
\end{equation}
Finally the EoS parameter of dark energy is obtained as
\begin{equation}\label{35}
w_{\phi}=-1+{\mathcal{A}(\mu,\nu,x)\over (x^2-1)^{3\over 2}},
\end{equation}
with
\begin{eqnarray}\label{36}
\mathcal{A}(\mu,\nu,x)=\big[\tilde{C}_1\left((\nu +{3\over 2})xP^{-\mu}_\nu(x)-(\nu+\mu+1)P^{-\mu}_{\nu+1}(x)\right)+\nonumber \\
\tilde{C}_2e^{-i\pi \mu}\left((\nu +{3\over 2})xQ^\mu_\nu(x)+(\mu-\nu-1)Q^{\nu+1}_\mu(x)\right)\big]^2.
\end{eqnarray}
Dimensionless constants $\tilde{C}_i$ are given by $\tilde{C}_i=\sqrt{9H_0^2\Omega_{\phi 0}^2\over 4V(\phi_c)\Omega_{m0}}C_i$. The slow roll condition  $\dot{\phi}^2\ll V$ is valid when $w_{\phi}=-1+{\dot{\phi}^2\over V(\phi)}\simeq -1$. In terms of  $\mathcal{A}$ (see \ref{35}), this can be rewritten as \begin{equation}\label{37}
\mathcal{A}(\mu,\nu,x)\ll (x^2-1)^{3\over 2}.
\end{equation}
This condition puts a constraint on the values of $\mu$ and $\nu$ in terms of the scale factor given by (\ref{28}) in the slow roll regime.  This constraint, as we will see, depends also on the value chosen for $a_c$. To see wether a chosen $\mu$ and $\nu$ satisfies (\ref{37}), one can employ numerical computation.

Since $\tilde{C}_i$ are not independent, the boundary condition
\begin{equation}\label{38}
w_{\phi}(a=a_c)=-1,
\end{equation}
gives $\mathcal{A}(\mu,\nu,a_c)=0$, leading to
\begin{equation}\label{39}
\tilde{C}_2=-{\frac {  \left( -\mu-\nu-1 \right) P^{-\mu}_{\nu+1}(x_c)
 +\left(\nu+3/2 \right)x_cP^{-\mu}_\nu(x_c)}{   \left( \mu-\nu-1 \right) Q^{\mu}_{\nu+1}(x_c) + \left( \nu+3/2 \right)x_cQ^{\mu}_{\nu}(x_c)   }}e^{i\pi \mu}\tilde{C}_1
\end{equation}
where $x_c=x(a_c)$.
One may introduce another relation between the parameters of the model
\begin{equation}\label{40}
w_{\phi}(a=1)=-1+\left({\Omega_{m0}\over \Omega_{\phi0}}\right)^{3\over 2}\mathcal{A}\left(\mu,\nu,\sqrt{1+{\Omega_{\phi 0}\over \Omega_{m0}}}\right)=w_0,
\end{equation}
where $w_{0}$ is the EoS parameter of dark energy at the present time.
Therefore if the parameters of the model are chosen according to (\ref{31}), (\ref{37}), (\ref{39}) and (\ref{40}), the EoS parameter of the quintessence is
determined by $w_\phi$ in (\ref{35}).

To obtain more specific relations for the parameters of the model, we may use astrophysical data. If based on Planck 2015 data \cite{Planck}, one chooses  $\Omega_{\phi_0}\simeq 0.685$, $\Omega_{m0}\simeq 0.315$,  and approximates $V(\phi_c)\approx \rho_{\phi0}$ (note that $\rho_{\phi0}$ has the same order as $\rho_{d0}$) , which is valid when $\dot{\phi}^2\ll V$ (or when $w_\phi\simeq -1$), (\ref{31}) becomes
\begin{eqnarray}\label{41}
\mu^2&=&{1\over 4}+ 1.41\left({V\mydprime(\phi_c)\over H_0^2}\right),\nonumber \\
\nu(\nu+1)&=&{3\over 4}-0.65 \left({V\mydprime(\phi_c)\over H_0^2}\right).
\end{eqnarray}
Similarly (\ref{40}) reduces to
\begin{equation}\label{42}
-1+0.312\mathcal{A}(\mu,\nu,1.78)=w_0,
\end{equation}
where $w_0=-1.006\pm 0.045 $ with $95\%$  C.L. \cite{Planck}.

\begin{figure}[H]
\centering\epsfig{file=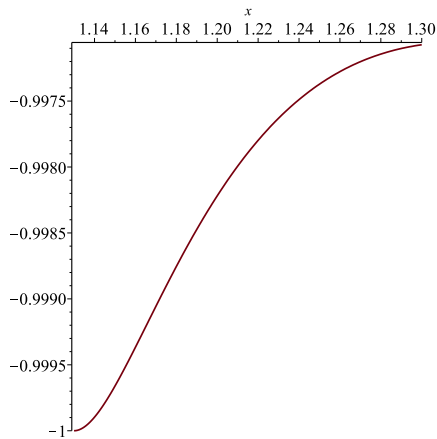,width=6cm,angle=0}
\caption{\footnotesize The EoS parameter of dark energy in thawing model in terms of $x$ for ${V\mydprime(\phi_c)\over H_0^2}=-10^{-4}$ and $a_c={1\over 2}$.} \label{fig1}
\end{figure}

\begin{figure}[H]
\centering\epsfig{file=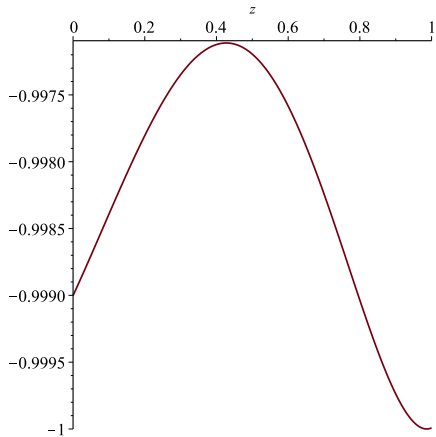,width=6cm,angle=0}
\caption{\footnotesize The EoS parameter of dark energy in thawing model in terms of the redshift $z$ for ${V\mydprime(\phi_c)\over H_0^2}=-10^{-4}$ and $a_c={1\over 2}$.} \label{fig2}
\end{figure}

Therefore, to study the evolution of $w_\phi$  one needs to specify ${V\mydprime(\phi_c)\over H_0^2}$ and $a_c$. As an illustration let us take a model with ${V\mydprime(\phi_c)\over H_0^2}=-10^{-4}$, e.g. for (\ref{13}) this is equivalent to $\mu=0.01H_0$. We assume that the symmetry breaking occurred at $a_c={1\over 2}$, and take $w_0=-0.99$ \cite{Planck}. The EoS parameter of dark energy is plotted numerically in terms of $x$ in figure \ref{fig1}. As it is clear from this figure ${\dot{\phi}^2\over V(\phi)}<0.01$ so ${A(\mu,\nu,x)\over (x^2-1)^3}={\dot{\phi}^2\over V(\phi)}\ll 1$.
Also, The EoS parameter of dark energy is plotted numerically in terms of the redshift $z={1\over a}-1$ in figure \ref{fig2}.

\section{Conclusions}
 The thawing model has been considered  in the literature previously \cite{chiba} without considering any interaction between dark sectors. The evolution of the quintessence near the maximum of its own potential (not the effective potential) was also discussed in the absence of interactions in \cite{scher} as the hilltop quintessence. However it was not emphasized why the quintessence is near its potential maximum in the present epoch.

In this paper, we considered the {\it coupled} quintessence in a thawing model and attributed the deviation of EoS parameter of dark energy from $w_\phi=-1$ and the dynamics of dark energy to the spontaneous $Z_2$ symmetry breaking of the effective potential. Indeed the beginning of the motion of the scalar field in the thawing model was related to the evolution of dark matter. We studied conditions required for the validity of this model for general potentials and couplings. We also investigated whether the scalar field can reside at the minimum of the effective potential and the stability of the model was discussed.

 Finally, in the last section, which includes our main results, we used the approximation $w_\phi\simeq -1$, and obtained analytical solutions for the quintessence and its EoS parameter near the extremum of the potential in terms of known functions.  As far as possible we determined the parameters of the model in terms of astrophysical data.

As an observational test and as an outlook, using equations (\ref{32}),  (\ref{34}) and (\ref{23}) and the Friedmann equation one can try to obtain an expression for the Hubble parameter in terms of the redshift parameter $z$. By deriving $H(z)$, one can obtain the luminosity distance redshift relation and compare it with type Ia supernova data \cite{supernova}.

\end{document}